\documentstyle[12pt,twoside,fleqn,espcrc1]{article}

\newcommand{\AmS}{{\protect\the\textfont2
  A\kern-.1667em\lower.5ex\hbox{M}\kern-.125emS}}

\title{Chaos and quantum interference effect in semiconductor ballistic
micro-structures}

\author{
	Shiro Kawabata\thanks{E-mail: shiro@etl.go.jp} \\
\bigskip
Physical Science Division,
	Electrotechnical Laboratory,\\
	1-1-4 Umezono, Tsukuba,  Ibaraki 305-8568,  Japan
\bigskip
}

\begin{document}
\maketitle

\begin{abstract}
We study the quantum-interference effect in the ballistic Aharonov-Bohm (AB)
billiard.
The wave-number averaged conductance
and the correlation function of the non-averaged conductance
are calculated by use of semiclassical theory.
Chaotic and regular AB billiards have turned
out to lead to qualitatively different semiclassical formulas for
the conductance with their behavior determined only by knowledge regarding
the underlying classical scattering.
\end{abstract}

\section{INTRODUCTION}

Recently a growing number of works accumulated around the subject of an
interplay between quantum chaos and ballistic quantum transport in
mesoscopic systems~\cite{rf:QCmeso,rf:Theory}.
Particularly, an important role of quantum chaos is addressed  in  quantum
interference in  Aharonov-Bohm (AB) billiards.
In this paper, we shall focus our attention to the
latest issue of our research on the
 $h/2e$ Altshuler-Aronov-Spivak (AAS)
oscillation~\cite{rf:KawabataAAS1,rf:KawabataAAS2}
and  $h/e$ AB oscillation~\cite{rf:KawabataAB}.
We shall present the analysis of
semiclassical theory of AAS and AB oscillation in an open chaotic system, e.g.,
the Sinai billiard.
Comparative study on other regular billiards will also be presented.

\section{ {\em h/2e} AAS OSCILLATION}

In this section, we shall consider an open single AB billiard in uniform
normal magnetic field $B$ penetrating $only$ through the hollow.
The ballistic weak localization (BWL) correction~\cite{rf:BJS} is most
easily discussed in terms of the reflection coefficient
$
R= \sum\nolimits_{n,m=1}^{N_M} \left| r_{n,m} \right|^2
$
, where $N_M$ and $r_{n,m}$ are the mode number and the reflection
amplitude, respectively.
Therefore, our starting point is the reflection amplitude~\cite{rf:Fisher}
\begin{equation}
r_{n,m}  =  \delta_{n,m} - i \hbar \sqrt{\upsilon_n \upsilon_m}
    \int dy \int dy' \psi_n^*(y') \psi_m(y)
          G(y',y,E_F),
  \label{eqn:a2}
\end{equation}
where $\upsilon_m(\upsilon_n)$ and $\psi_n(\psi_m)$ are the longitudinal
velocity
and transverse wave function for the mode $m (n)$ at a pair of lead wires
attached
to the dot. $G$ is the retarded Green's function
connecting points $(x,y)$ and $(x',y')$ on the left and right leads,
respectively.

In order to carry out the semi-classical approximation, we replace $G$ by
its semiclassical Feynman path-integral expression~\cite{rf:SC},
\begin{equation}
G^{sc}(y',y,E) = \frac {2 \pi} {(2 \pi i \hbar)^{3/2}} \sum_{s(y,y')}
  \sqrt{D_s}
 \exp \left[
                            \frac i {\hbar} S_s (y',y,E) - i \frac \pi {2}
\mu_s
                       \right]
,
  \label{eqn:a3}
\end{equation}
where $S_s$ is the action integral along classical path $s$,
 \( D_s = ( \upsilon_F \cos{\theta'})^{-1} \left| ( \partial \theta
/\partial y'  )_y \right| \)
, \( \theta \) (\( \theta' \)) is the incoming (outgoing) angle, and \(
\mu_s \) is the Maslov index.

Substituting eq.(\ref{eqn:a3}) into eq.(\ref{eqn:a2}) and
carrying out the double integrals by the saddle-point approximation,
we obtain
\begin{equation}
r_{n,m} = - \frac {\sqrt{2 \pi i \hbar}} {2 W} \sum_{s(\bar n,\bar m)}
  {\rm sgn} (\bar n) {\rm sgn} (\bar m) \sqrt{\tilde D_s}
\exp{
        \left[
              \frac i {\hbar} \tilde S_s (\bar n,\bar m;E)-i \frac \pi {2}
\tilde \mu_s
        \right]
	  },
  \label{eqn:a4}
\end{equation}
where $W$ is the width of leads and \( \bar m = \pm m \). The summation
is over trajectories between the cross sections at $x$ and $x'$ with angles
\( \sin{\theta} =  \bar{m} \pi / k W  \)
and
\( \sin{\theta'} =  \bar{n} \pi / k W  \).
In eq.(\ref{eqn:a4}),
\(
  \tilde{S_s} (\bar{n},\bar{m};E) = S_s(y'_0,y_0;E)+ \hbar \pi ( \bar{m} y_0
- \bar{n} y'_0 ) / W
\)
,
\(
  \tilde{D_s} = ( m_e \upsilon_F \cos{\theta'})^{-1} \left| ( \partial y
/\partial \theta' )_{\theta} \right|
\)
and
\(
\tilde{\mu_s} = \mu_s + H \left( -( \partial \theta / \partial y )_{y'} \right)
                      + H \left( -( \partial \theta' / \partial y'
)_{\theta} \right),
\)
respectively, where $H$ is the Heaviside step function.

Taking the diagonal approximation, there is a natural procedure for finding
the average of
$
\delta R_D = \sum\nolimits_{n=1}^{N_M}\delta R_{n,n}
$
over wave-number $k$, which is denoted by $\delta {\cal R}_D$.
Therefore the contribution to the BWL correction term $\delta {\cal R}_D
(\phi)$
is just given by
the pair of time reversal paths.
With use of the extended semiclassical theory~\cite{rf:Takane}, we can take
account of the
offdiagonal part and the influence of the small-angle diffraction as
\( \delta {\cal R} (\phi) =   \delta {\cal R}_D (\phi)/2 \),
for the case that the width of the lead wires are equal.
Then we obtain the full quantum correction of the conductance for chaotic AB
billiards as
\begin{equation}
    \delta g (\phi) =
				 -
				 \frac{e^2}{\pi \hbar}
				 \frac1{4}
				 \frac{\cosh \eta -1}{\sinh \eta}
				      \left\{
					        1 +
							2 \sum_{n=1}^{\infty}
							 \exp  \left( -
\eta n \right)
					          \cos \left(  4 \pi n
\frac{\phi}{\phi_0} \right)
					  \right\}
 ,
  \label{eqn:a6}
\end{equation}
where $\eta = \sqrt{2 T_0 \gamma /\alpha}$.~\cite{rf:KawabataAAS1}
System-dependent constants $\alpha$, $T_0$ and $\gamma$ correspond to the
variance of the winding number distribution~\cite{rf:Berry}, the dwelling
time for the shortest classical
orbit and the escape rate, i.e., $N(T) \sim \exp (-\gamma
T)$,~\cite{rf:BJS,rf:LDJ} respectively.
In eq. (\ref{eqn:a6}), the period of the conductance oscillation is $h/2e$,
analogous to the AAS oscillation in diffusive systems.~\cite{rf:AAS}
In chaotic case, the oscillation amplitude decays
exponentially with increasing the rank of higher harmonics $n$, so that the
main contribution
to the conductance oscillation
comes from $n=1$ component which oscillates with the period of $h$/2$e$.

On the other hand, for regular AB billiards,
we obtain
\begin{equation}
\delta g (\phi)  = - \left| \delta g (0) \right|
                     \frac{
					        \displaystyle{
					        1 +
							2 \sum_{n=1}^{\infty}
							  F  \left(
\beta-\frac1{2}, \beta+\frac1{2};-\frac{n^2}{2\alpha}  \right)
					          \cos \left(  4
\pi n \frac{\phi}{\phi_0} \right)
					         }
						   }
						   {
						    \displaystyle{
					        1 +
							2 \sum_{n=1}^{\infty}
							  F  \left(
\beta-\frac1{2}, \beta+\frac1{2};-\frac{n^2}{2\alpha}  \right)
						     }
						   }
   ,
  \label{eqn:a7}
\end{equation}
where
$
F
$
and $z$ are the hypergeometric function of confluent type and the exponent of
dwelling time distribution \( N(T)  \sim T^{- z} \)~\cite{rf:BJS,rf:LDJ},
respectively.
In eq.(\ref{eqn:a7}) the oscillation amplitude decays algebraically for
large $n$,
therefore the higher-harmonics components give noticeable contribution to
magneto-conductance oscillation.
These discoveries indicate that {\em the $h/2e$ AAS oscillation occurs in
both ballistic and diffusive
systems forming AB geometry and the behavior of higher harmonics components
reflects a difference between chaotic and non-chaotic classical dynamics}.

In real experiments~\cite{rf:Taylor}, magnetic field would be applied
to $all$ region (both the hollow and annulus) in the billiard.
In this situation, we will observe $h/2e$ oscillation together with  the
negative magneto-resistance and dampening of the oscillation amplitude with
increasing magnetic field as in the case of diffusive AB
rings.~\cite{rf:KawabataAAS3}

\section{ {\em h/e} AB OSCILLATION}
In previous section, we have investigated the $h/2e$ AAS oscillation for
$energy-averaged$ magneto-conductance.
The result of quantum-mechanical calculations~\cite{rf:KawabataAAS2}
indicated that the period of the energy averaged conductance,
\begin{equation}
     <g(\phi)>_E =  \frac{1}{\Delta E} \int _{E_F -\Delta E /2} ^{E_F +
\Delta E /2} g(E,\phi) dE
	 ,
\end{equation}
changed from $h/2e$
to $h/e$, when the range of energy average $\Delta E$ is decreased.
In this section, we shall calculate the correlation function $C(\Delta
\phi)$ of the $non-averaged$ conductance
by using the semiclassical theory
and show that $C(\Delta \phi)$ is qualitatively different between
chaotic and regular AB billiards.~\cite{rf:KawabataAB}

The fluctuations of the conductance
$g=(e^2/\pi \hbar) T(k) = (e^2/\pi \hbar) \sum \nolimits _{n,m} \left|
t_{n,m} \right|^2 $
are defined by their deviation from the classical
value;
$ \delta g \equiv g - g_{cl} $
.
In this equation $g_{cl} = (e^2/\pi \hbar) T_{cl}$, where $T_{cl}$ is the
classical total transmitted intensity.
In order to characterize the $h/e$ AB oscillation, we define the correlation
function of the oscillation in
magnetic field $\phi$ by the average over $\phi$,
$
 C(\Delta \phi) \equiv \left< \delta g(\phi) \delta g(\phi+\Delta
\phi)\right>_{\phi}
$ .
With use of the ergodic hypothesis, $\phi$ averaging can be replaced by the
$k$ averaging, i.e.,
$
 C(\Delta \phi) = \left< \delta g(k,\phi) \delta g(k,\phi+\Delta \phi)
\right>_{k}
$ .
The semiclassical correlation function of conductance is given by
%
%
%
%
\begin{eqnarray}
C (\Delta \phi)
             =
				  \left( \frac{e^2}{\pi \hbar} \right)^2
                  \frac{1}{8}
				  \left(
				  \frac{\cosh \eta \!-\! 1}{\sinh \eta}
				  \right)^2
				  \cos \left( 2 \pi \frac{\Delta
\phi}{\phi_0} \right)
                  \left\{
                     1 + 2 \sum_{n=1}^{\infty} e^{-\eta n}
			         \cos \left( 2 \pi n \frac{\Delta
\phi} {\phi_0} \right)
                  \right\}^2
				  ,
\label{eqn:e6-2-24}
\end{eqnarray}
%
%
where \( \eta = \sqrt { 2 T_0 \gamma / \alpha } \).
In deriving eq.~(\ref{eqn:e6-2-24}) we have used the exponential dwelling time
distribution and the Gaussian winding number distribution.

On the other hand, for the regular cases, we use $N(T) \sim
T^{-\beta}$.~\cite{rf:BJS}
Assuming as well the Gaussian winding number distribution , we get
%
%
\begin{eqnarray}
 C (\Delta \phi) =
                  C (0)
 				  \cos \left( 2 \pi \frac{\Delta
\phi}{\phi_0} \right)
				  \left\{
                     \frac{
                     1+2 \displaystyle{\sum_{n=1}^{\infty}
                      F  \left( \beta-\frac1{2},
\beta+\frac1{2};-\frac{n^2}{2\alpha}  \right)
                      \cos \left( 2 \pi n \frac{\Delta \phi} {\phi_0} \right) }
					  }
					  {
                     1+2  \displaystyle{\sum_{n=1}^{\infty}
                     F  \left( \beta-\frac1{2},
\beta+\frac1{2};-\frac{n^2}{2\alpha}  \right)}
 					  }
				  \right\}^2
				  .
\label{eqn:e6-2-29}
\end{eqnarray}
%
%
Therefore, these results indicate that the difference of $C(\Delta \phi)$ of
these ballistic AB billiards
can be attributed to the difference between chaotic
and regular classical scattering dynamics.

\section{SUMMARY}

The statistical aspect of classical open AB billiards is characterized by the
dwelling time distributions $N(T)$ which reflects the integrability of
the system. For chaotic billiards, $N(T)$ obeys the exponential
distribution. On the other hand, for regular AB billiards, $N(T)$
obeys the power-law distribution. The difference of the
distributions affects the AAS and AB oscillation in the ballistic regimes.

In conclusion, we indicated a way to observe a quantum signature of
chaos through the novel quantum interference phenomena in ballistic AB
billiards.

I would like to acknowledge
Y. Takane, Y. Ochiai, H.A. Weidenm\"uller, R.P. Taylor, F. Nihey and K.
Nakamura
for valuable discussions and comments.

\end{document}